\documentclass[%
 reprint,
 showpacs,
 showkeys,
 nofootinbib,
 amsmath,amssymb,
 pra,
 floatfix,
 ]{revtex4-1}

\usepackage{graphicx}
\usepackage{epstopdf}
\usepackage{eepic}
\usepackage{color}
\usepackage{amsmath}
\usepackage{amsthm}
\usepackage{comment}
\usepackage{multirow}

\usepackage{amsmath,amssymb,amsfonts,url,amsthm,enumerate}
\usepackage{graphicx,tikz,pgf,epsf}
\usetikzlibrary{backgrounds,fit,decorations.pathreplacing,shapes.misc,calc}  
\tikzset{cross/.style={cross out, draw=black, minimum size=2*(#1-\pgflinewidth), inner sep=0pt, outer sep=0pt},
cross/.default={1pt}}

\usepackage{algorithm,algcompatible}
\algnewcommand\algorithmicreturn{\textbf{return}}
\algnewcommand\RETURN{\State \algorithmicreturn}%

\usepackage{dcolumn}
\usepackage{bm}

\theoremstyle{plain}

\usepackage{array}
\newcolumntype{H}{>{\setbox0=\hbox\bgroup}c<{\egroup}@{}}

\newcommand{\ket}[1]{\left|#1\right\rangle}
\newcommand{\bra}[1]{\left\langle #1\right|}
\newcommand{\braket}[2]{\left\langle #1 \right.\left| #2 \right\rangle}

\newcommand{\0}{{\bf 0}}
\newcommand{\1}{{\bf 1}}
\newcommand{\?}{{\bf ?}}
\newcommand{\p}{\mathbf{p}}

\begin{document}
\bibliographystyle{plainnat} 


\title{Diagnosis of single faults in quantum circuits}


\author{Debajyoti Bera}
\email{dbera@iiitd.ac.in}
\affiliation{Indraprastha Institute of Information Technology,
Okhla Industrial Estate Ph-III, New Delhi, India 110020}
\author{Subhamoy Maitra}
\email{subho@isical.ac.in}
\affiliation{Indian Statistical Institute, 203 B T Road, Kolkata 700108, India}
\author{Sparsa Roychowdhury}
\email{sparsa.roychowdhury@gmail.com}
\affiliation{Indian Statistical Institute, 203 B T Road, Kolkata 700108, India}
\author{Susanta Chakraborty}
\email{susanta.chak@gmail.com}
\affiliation{Indian Institute of Engineering Science and Technology, Shibpur, Howrah 711103, India}

\date{\today}

\begin{abstract}
Detecting and isolating faults is crucial for synthesis of quantum circuits. Under the single
fault assumption that is now routinely accepted in circuit fault analysis, we 
show that the behaviour of faulty quantum circuits can be fully characterized by
the single faulty gate and the corresponding fault model. This allows us to
efficiently determine test input states as well as measurement
strategy that can be used to detect every single-gate fault using very few test
cases and with minimal probability of error; in fact we demonstrate that most
commonly used quantum gates can be isolated without any error under the single
missing gate fault (SMGF) model. We crucially exploit the quantum nature of circuits to show vast
improvement upon the existing works of
automatic test pattern generation (ATPG) for quantum circuits.
\end{abstract}


\maketitle





\section{Introduction}
\label{intro}
Detection and diagnosis of faults in classical digital circuits have been part of
mainstream circuit manufacturing research and industry for several decades.
A common approach for this is to analyze outputs when a circuit is given a fixed set of
carefully chosen inputs (also known as patterns). ATPG (automated test pattern generation)
techniques essentially try to efficiently generate an effective set
of such inputs. This is computationally challenging because it belongs to the
category of NP-complete problems~\cite{IS75}.
However, extremely efficient heuristics have led to successful adoption of ATPG
in VLSI.

ATPG is an obvious avenue to explore fault detection in quantum circuits; however, current results on
ATPG for quantum circuits are few and do not seem to fully exploit the power of
quantum computation in generating an efficient set of test patterns.
In this regard, we noted the idea of quantum tomographic testing that has 
been discussed earlier in~\cite{paler1,paler2}.
Like quantum state tomography, quantum tomographic testing requires
multiple measurements on the circuit in question on various inputs; the
histogram or the frequency counts of output patterns can be used to approximately determine the
output states. The pairs of input and output states are then analyzed to detect
and diagnose faults in the circuit.

Theoretically the number of possible faults is endless for any quantum circuit. Practically 
however, a method of synthesizing a circuit limits the possible set of faults.
{\em Single fault assumption} is now routine used in academia and industry in
which the cause of a circuit failure is attributed to only one faulty component
(gate).
Hayes et al.\ reported that the commonly used ``stuck-at'' fault-model and ``bridging
fault-model are not very apt for quantum circuits~\cite{HayesPolian}. 
They proposed the missing gate fault (MGF) model 
in which one or more quantum gates are missing, i.e., these gates behave like an identity operator. 
ATPG for quantum
circuits have since then largely focused around {\em single MGF} (SMGF). It is of course
clear that if a gate behaves almost like an identity operator, then detecting
whether it is present or missing is going to be difficult, if not impossible.
However, formal understanding of this statement was not available so far.

In this work we explain what it means for a gate to be hard to diagnose
(maybe more computation is required but no faulty gate is impossible to diagnose). We further show how to efficiently obtain a set of input states and measurement operators to cover all faulty gates with much less trials 
compared to existing results~\cite{paler1,paler2}.
Unlike these works, our detailed explanations pertain to
arbitrary single gate fault models, including SMGF;
the faults may even be different for each gate. Furthermore, we do not
require any internal modification of the circuit (unlike \cite{paler2}).

The technical contribution
is essentially answers to a set of questions raised, but left unaddressed, in
\cite{paler2}. Central to these are three observations, all of which are peculiarities of quantum
circuits not present in their classical cousins. Suppose we have two states
$\ket{u}$ and $\ket{v}$ whose output, when measured identically, generate the
probability distributions $\{u_m\}$ and $\{v_m\}$, respectively, on measurement
outcomes.
The first observation is that
the ``difference'' caused by faulty behaviour of any gate is preserved subsequently. In other words,
statistical distance between $\{u_m\}$ and $\{v_m\}$ 
does not change if an identical quantum circuit is applied on
both the states before measuring.

The second observation is that using the same measurement operators on some
combination of the input states ($\alpha\ket{u} + \beta\ket{v}$) generates a
probability distribution which is {\em not a linear combination}
($\{|\alpha|^2 u_m + |\beta|^2 v_m\}$) of the earlier probability distributions.
Therefore, it makes sense to explicitly consider input states that are in
superpositions of standard basis states.

Finally, motivated by the last reason, we ought to take advantage of the
generalized measurement operators that quantum systems allow beyond the usual
practice of measuring {\em in the standard basis}.
It is therefore immediate that, unlike earlier approaches
(\cite{paler1,paler2}), we ought to be looking for input states
that could be super-position of basis states, and measurement operators more
general than projective measurements in the standard basis.

We will denote by $C$ the circuit to be diagnosed which acts on $n$ qubits and
represent its gates when enumerated in the standard
manner, by $G_1, G_2. \ldots G_s$. To simplify our explanation, we will say
$G_0$ is faulty to mean that $C$ is fault-free.
In the fault model we consider, at most one
of these $s$ gates is faulty, and when faulty, the corresponding faulty
behaviour (operator) is known to us. The operators for the
fault-free and faulty $i$-th gate are denoted by $G^i$ and $G_f^i$,
respectively ($G_f^i$ is set to the identity operator under the
SMGF model).
Let $C^0$ denote a circuit in which no gate is faulty, and $C^i$ denote a
circuit in which (only) the $i$-th gate is faulty. That is, $C^0=G^s \ldots
G^{i+1} G^i G^{i-1} \ldots G^1$ and $C^i=G^s \ldots G^{i+1} G^i_f G^{i-1} \ldots$.
We want to {\em detect} if $C$ is fault-free or
faulty, and furthermore, if faulty then {\em diagnose} the offending gate; in other
words, we want to know $C=C^j$ for which $j \in \{0, \ldots s\}$. We illustrate
the effectiveness of our strategy by applying it on a benchmark circuit
``3qubitcnot''.

\section{Detecting if a specific gate is faulty}
\label{sec:algo-faulty-gate}
This section contains the main tools of this paper. Suppose we are
told that {\em all but the $i$-th gate
are fault-free}. We want to detect if the $i$-th gate is fault-free or faulty.
For the sake of brevity, we will use the following notation in this
section: $G=G^i$, $G_f=G^i_f$, $C=C^0$ and $C'=C^i$. Thus, we want to know if the
$i$th-gate is $G$ or $G_f$.

\paragraph*{} The high-level idea of our approach is to
\begin{enumerate}
\item Find an input state $\ket{\phi}$ such that $\ket{\psi}=C\ket{\phi}$ and $\ket{\psi'}=C'\ket{\phi}$ are at the farthest
``distance'' possible.
\item Find measurement operators which can distinguish between those two
    faraway states $\ket{\psi}$ and $\ket{psi'}$ with minimal probability of failure.
\end{enumerate}

The fact that these can be done in general is well-known~\cite{helstrom}. We
reformulate the known results and describe how to derive optimum input state and measurement
operators to solve the problem described above.

\subsection{Optimal input state}
\label{subseq:optimal-state-faulty-gate}
The appropriate measure of ``distance'' for pure quantum states with respect to
distinguishability is the trace distance defined by
$\displaystyle D(\ket{\psi}, \ket{\psi'}) =
\sqrt{1-|\braket{\psi}{\psi'}|^2}$.
Trace distance is also equal to the maximum
L1 distance of the probability distributions obtained from the two states upon
measurement~\cite{nc}.
Given two operators $G$ and $G_f$, we say that a state $\ket{\phi}$ is
a $(G,G_f)$-separator (similarly, $(C,C')$-separator) if this state, given as
input, maximizes the trace distance between
$G\ket{\phi}$ and $G_f\ket{\phi}$ (respectively, $C\ket{\phi}$ and $C'\ket{\phi}$).

Therefore, our immediate goal is to find a $(C,C')$-separator input
state $\ket{\phi}$ which minimizes $|\braket{\psi}{\psi'}|$.
Our main observation here is that we can decompose our circuits into common sub-circuits excluding the
$i$-th gate: $C=C_2 G C_1$ and $C' = C_2 G_f C_1$. Let $S=G^\dagger G_f$.
Without loss of generality, we can consider that $G$ (hence, $G'$ and $S$) acts on all $n$ qubits
(possibly by tensoring with an identity operator of suitable dimensions).

Let the the eigenvalues of $S$ be denoted by $e^{-i\theta_1} \ldots e^{-i\theta_m}$ (including
duplicates) and the corresponding eigenvectors by $\ket{v_1}, \ldots \ket{v_m}$. Let $\bar{a}=\{a_1 \ldots a_m\}$ 
be a solution to this optimization problem:
\begin{align}
    \label{eqn:1}
    \mathbf{OPT:~~~} \min & \sum_{j} a_j^2 + \sum_{j\not=k} a_j a_k \cos(\theta_j -
    \theta_k) \\
	 & \mbox{ where}
	\sum_j a_j = 1,~~0 \le a_j \le 1\nonumber
\end{align}

First observe that minimizing Eqn.\ \ref{eqn:1} is equivalent to minimizing
    $\displaystyle\sqrt{ \sum_{j} a_j^2 + \sum_{j\not=k} a_j a_k \cos(\theta_j - \theta_k)
}$
\begin{align*}
    & = \Big| \sum_j a_j\cos\theta_j - i\sum_j a_j\sin\theta_j
\Big|\\
    & = \Big| \sum_j a_j e^{-i\theta_j} \Big| = |\bra{\phi'}S\ket{\phi'}|
\end{align*}
where, $\ket{\phi'} = \sum_j \sqrt{a_j} \ket{v_j}$ is a state on $n$ qubits.

Therefore the optimum $\bar{a}$ for $\mathbf{OPT}$ minimizes
$|\bra{\phi'}G^\dagger G_f \ket{\phi'}|$, which makes $\ket{\phi'}$ a $(G,G_f)$-separator.
We can now choose $\ket{\phi}=C_1^\dagger \ket{\phi'}$ as our required $(C,C')$-separator input. Since
${|\bra{\phi'}S\ket{\phi'}|}  = {|\bra{\phi}C_1^\dagger G^\dagger C_2^\dagger C_2
G_f C_1 \ket{\phi}|}
= {|\bra{\phi}C^\dagger C'\ket{\phi}|} 
= {|\braket{\psi}{\psi'}|}$, the optimum $\bar{a}$ also minimizes
$|\braket{\psi}{\psi'}|$ and this minimum value is simply $|\sum_j a_j
e^{-i\theta_j}|$.

The above optimization problem is not a computational hurdle for three reasons.
First, the number of variables is exponential only in the dimension of the
gate involved, which is usually quite small in practice. Secondly, {\bf OPT} has
the form of an  equality-constrained quadratic program for which efficient
algorithms exist.

The final reason is the interesting fact that the separator input for a gate in a circuit
depends fundamentally on the gate in question and corresponding fault model. It does not depend at all on the
portion of the circuit coming after the faulty gate ($C_2$), and its dependence on the
portion of the circuit before the faulty gate ($C_1$)
is really incidental. Therefore, it is feasible to
have a pre-computed table of $(G,G_f)$-separators for different gates under common fault models. The required
separator input for any circuit can be obtained
by running the first portion of the circuit in reverse on a gate-separator input. 
Quantum circuits are usually built using a small set
of gates that operate on a small number of qubits. Therefore, the major
computation tasks of eigen-decomposition of $S$ and solving $\mathbf{OPT}$ can be
done only once and reused as and when needed.

If $G$ acts on $n'$ qubits and $n' \ll n$ (say, $n'=1$ or 2), then it is
possible to solve {\bf OPT} using the  larger $n$-qubit operator
$I_{n-n'} \otimes G_i$. This may be computationally expensive, so a
better
alternative is to let $S=G^\dagger G_f$ as before, and let $T=I_{n-n'} \otimes S$ be the extension
of $S$ to $n$ qubits. If $\{(e^{-i\theta_j}, \ket{v_j})\}$ are the eigenpairs of $S$ then
it is easy to see that $\{(e^{-i\theta_j}, \ket{v_j} \otimes \ket{0}^{\otimes
n-n'})\}$ are the eigenpairs of $T$.
Thus our required input state can be derived as $\ket{\phi} =  C_1^\dagger \big(\ket{\phi'} \otimes
\ket{0}^{\otimes (n-n')} \big)$ where $\ket{\phi'}$ is a $(G,G_f)$-separator
input state. For example, if $G$ is a single qubit gate, then we only need to
store that $\ket{\phi'}$
is $\frac{1}{\sqrt{2}}\left( \ket{v_1} + \ket{v_2} \right)$ where
$\ket{v_1}$ and $\ket{v_2}$ are the eigenvectors of $G^\dagger G_f$,
irrespective of the value of $n$.

If the fault in question belongs to the single missing gate fault model, then we
can treat it is a special case of the above where $G_f=I$ and therefore
$S=G^\dagger$. Table \ref{table:gate-separator-input}
presents the separator input states for various commonly used gates in the
SMGF model.

\begin{table}
\resizebox{0.8\columnwidth}{!}{%
\begin{tabular}{|l|c|c|}
\hline
Gate & Separator input states & Error prob.\\
\hline
    Hadamard &  
    $\left[
        \begin{array}{c}
	    -\sqrt{\frac{\sqrt{2}-1}{2\sqrt{2}}} \quad \\[1em]
	    -\frac{\sqrt{1-\frac{1}{\sqrt{2}}}}{2-\sqrt{2}}
        \end{array}
    \right]$ &
        0.00 \\
\hline
    Phase &
    $\left[
	\begin{array}{c}
	    1/\sqrt{2} \\
	    1/\sqrt{2}
	\end{array}
    \right]$
    &
    0.15\\
\hline
    {CNOT} &
    
	$\left[
	    \begin{array}{c}
		0.4082\\
		0.4082\\
		-0.2113\\
		0.7887
	    \end{array}
	\right]$
     &
     0.00\\
\hline

    $R_y(\pi/6)$&
    $\left[
	\begin{array}{c} 
	    0 \\
	    -1 
	\end{array}
    \right] $
    &
   0.37 \\
    
\hline
    $R_z(\pi/16)$ &
    $\left[
	\begin{array}{c}
	    1/\sqrt{2} \\
	    1/\sqrt{2} 
	\end{array}
    \right] $
    &
    0.45
\\
    
\hline
    Toffoli &
    
	$\left[
	    \begin{array}{c}
		 0.2673\\
		 0.2673\\
		 0.2673\\
		 0.2673\\
		 0.2673\\
		 0.2673\\
		-0.3110\\
		 0.6890
	    \end{array}
	\right]$
     &
     0.00\\
     
\hline
    Pauli-X &
    $\left[
	\begin{array}{c}
	    0 \\
	    1 
	\end{array}
    \right] $ 
    &
    0.00\\
\hline
    
    Pauli-Y &
    $\left[
	\begin{array}{c}
	   -i \\
	    0 
	\end{array}
    \right] $ 
    &
    0.00 \\
\hline

    Pauli-Z &
    $\left[
	\begin{array}{c}
	    1/\sqrt{2} \\
	    1/\sqrt{2}
	\end{array}
    \right] $
    &
     0.00 \\
\hline
\end{tabular}
}
\caption{\label{table:gate-separator-input}Optimal $(G,I)$-separator inputs
    states ($\ket{\phi'}$) and corresponding probability of error ($\delta$) using optimal projective
    measurement operators for detecting missing quantum gates. The states are
    described as vectors in the standard basis.}
\end{table}

It should be obvious that our method of
decomposing a circuit into portions before and after the gate in question can
also be used for multiple missing/defective gate faults as long as the faulty
gates can be grouped together and the circuit can be sliced around them. For
example, our method is applicable to multiple gate faults if they act on
distinct set of qubits, and/or are adjacent to each other; trivial extension is
required to the computation of optimal state described earlier.

\subsection{Optimal measurement}
\label{subseq:optimal-measurement-faulty-gate}
Once we have obtained the optimal input state $\ket{\phi}$, we can compute the
two possible output states $\ket{\psi}=C\ket{\phi}$ and $\ket{\psi'}=C'\ket{\phi}$.
Quantum states are manifested only by their measurement outputs. It
is thus important to design and implement 
measurement operators that are able to
distinguish between these states. 
and thereby determine if the circuit in question is $C$ or $C'$.
However, unlike input states, measurement operators depend on the
actual circuit and has to be computed once for every circuit and every fault
model.

The question of distinguishing between two given quantum states is one of
the classical problems of quantum computing\cite{chefles}. Two states can be differentiated
(using measurements) with certainty if and only if they are orthogonal. So, if
$G_f$ is almost same as $G$, then obviously 
no measurement should be able to distinguish between them 
with high confidence. 
We show below that we can distinguish with high confidence for gates with low
value of $|\sum_j a_je^{-i\theta_j}|$ (obtained by solving {\bf OPT}).

\begin{figure}[t]
\resizebox{0.7\columnwidth}{!}{%
    \begin{tikzpicture}[thick]
    \coordinate[] (A) at (0,0);
    \coordinate[label=below right:{\Large $\ket{\psi'}$}] (P2) at (5,2.5);
    \coordinate[label=above left:{\Large $\ket{\psi}$}] (P1) at (2,4.17);
    \coordinate[label=above left:{\Large $\ket{\omega_+}$}] (W) at (0,4.4);
    \coordinate[label=below right:{\Large $\ket{\omega_-}$}] (W2) at (6,0);

\draw[thick,->] (A) -- (P1);
\draw[thick,->] (A) -- (P2);
\draw[thick,->] (A) -- (W);
\draw[thick,->] (A) -- (W2);

\begin{scope}
\path[clip] (A) -- (P2) -- (P1);
\draw[thick] (A) circle (10mm);
\node at ($(A)+(40:15mm)$) {\large $ke^{i\kappa}$};
\end{scope}
\begin{scope}
\path[clip] (A) -- (P2) -- (W);
\draw[thick] (A) circle (27mm);
\node at ($(A)+(45:30mm)$) {\large $\delta$};
\end{scope}
\begin{scope}
\path[clip] (A) -- (P1) -- (W2);
\draw[thick] (A) circle (20mm);

\node at ($(A)+(40:23mm)$) {\large $\delta$};
\end{scope}

\end{tikzpicture}
}
\caption{\label{fig:angle} Schematic diagram for the Helstrom projective
measurement basis. The angles represent the inner product between the
corresponding state vectors.}
\end{figure}
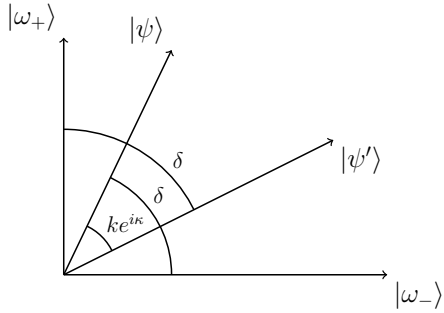

There are two known modes of distinguishing between a pair of states. Helstrom
measurement is a two-output (von Neumann) projective measurement which {\em
minimizes} the error of incorrect labelling\cite{helstrom}. 
If we prohibit incorrect outcome and instead allow our measurement operators to either label a state
with certainty or report ``\?''{\em (inconclusive)}, then we would be performing
what is known as {\em unambiguous state discrimination} (USD).
USD is commonly
achieved by employing a POVM\cite{IDPi,IDPd,IDPp}, a generalized measurement
operator. 
We will use Helstrom projective measurement in the rest of this paper
for explaining our technique; however, we can also use USD for doing the same
and even combine both of these techniques for different gates and faults.

\begin{table*}[t]
\begin{tabular}{|l|l|l|l|l|l|l|l|}
\hline
Test\textbackslash Circuit & $C^0$ & $C^1$ & $C^2$ & $C^3$ & $C^4$ & $C^5$ \rule{0pt}{1em} &
$C^6$ \\
\hline
Test(F1) & (1.00,0.00,0.00) & (0.00,1.00,0.00) &
(0.00,0.56,0.44)&
(0.07,0.50,0.43)&
(0.94,0.01,0.05) &
(0.98,0.01,0.01)&
(0.38,0.13,0.49)\\
\hline
Test(F2) & 
(1.00,0.00,0.00)&
(0.73,0.12,0.15)&
(0.00,1.00,0.00)&
(0.50,0.00,0.50)&
(0.87,0.00,0.13)&
(0.99,0.00,0.01)&
(0.76,0.00,0.24)\\
\hline
Test(F3) & 
(1.00,0.00,0.00)&
(1.00,0.00,0.00)&
(0.50,0.00,0.50)&
(0.00,1.00,0.00)&
(0.87,0.06,0.07)&
(0.98,0.00,0.02)&
(0.86,0.00,0.14)\\
\hline
Test(F4) &
(0.75,0.25,0.00)&
(0.75,0.25,0.00)&
(0.37,0.13,0.50)&
(0.00,0.00,1.00)&
(0.25,0.75,0.00)&
(0.74,0.25,0.01)&
(0.19,0.07,0.74)\\
\hline
Test(F5) &
(0.60,0.40,0.00)&
(0.40,0.27,0.33)&
(0.20,0.30,0.50)&
(0.55,0.38,0.07)&
(0.52,0.35,0.13)&
(0.40,0.60,0.00)&
(0.25,0.25,0.50)\\
\hline
Test(F6) &
(1.00,0.00,0.00)&
(0.56,0.08,0.36)&
(0.06,0.22,0.72)&
(0.10,0.42,0.48)&
(0.87,0.02,0.11)&
(0.99,0.01,0.00)&
(0.00,1.00,0.00)\\
\hline
\end{tabular}
\caption{\label{table:fault-table-proj}
Diagnostic fault table for circuit 3qubitcnot with single missing gate faults
using Helstrom projective measurements}
\end{table*}

For Helstrom projective measurement, we want to create an orthonormal basis $\ket{\omega_+}$ and
$\ket{\omega_-}$ which spans $\ket{\psi}$ and $\ket{\psi'}$. This basis will be
used for measurement and we will infer the state as $\ket{\psi}$ or $\ket{\psi'}$
upon measurement outcome $\ket{\omega_+}$ or $\ket{\omega_-}$, respectively.
We want to minimize the probability of error (when state is $\ket{\psi}$ but outcome is
incorrectly $\ket{\omega_-}$ and similarly for the other pair); so the basis
states should be maximally away from the output states, i.e.,
$|\braket{\omega_-}{\psi}|^2 = |\braket{\omega_+}{\psi'}|^2$. We denote
the corresponding probability of error by $\delta$.

We will represent by $k e^{i\kappa}$ the complex number $\braket{\psi}{\psi'}=\sum_j
a_je^{-i\theta_j}$ in which $a_j$'s are the
solution of $\mathbf{OPT}$ and $e^{i\theta_j}$ are the eigenvalues of $S =
G^\dagger G_f$. We first represent our states in terms of our basis states,
i.e., $\ket{\psi} = \alpha_1
\ket{\omega_+} + \beta_1 \ket{\omega_-}$ and $\ket{\psi'} = \alpha_2
\ket{\omega_+} + \beta_2 \ket{\omega_-}$. Without loss of generality, we can take $\alpha_1$
as a real number $r_1$. The condition of equal probability of error enforces these
representations: $\beta_1=r_2 e^{ix_1}$ for some real $r_2=\sqrt{1-r_1^2}$, $\alpha_2 =
r_2 e^{ix_2}$ and $\beta_2 = r_1 e^{ix_3}$. The inner product
$\braket{\psi}{\psi'}$ then simplifies to $r_1 r_2 e^{-ix_2}(1 +
e^{i(x_3+x_2-x_1)})$ which we need to equate to $k e^{i\kappa}$. One possible
solution is given by: $x_1=0$, $x_2=-\kappa$, $x_3=\kappa$ and $r_{1,2} = \left(
\sqrt{1+k} \pm \sqrt{1-k} \right)/2$ which produces this basis.
\begin{align*}
    \ket{\omega_+} & = \frac{r_1}{r_1^2 - r_2^2 e^{-2i\kappa}} \ket{\psi} -
    \frac{r_2e^{-i\kappa}}{r_1^2 - r_2^2 e^{-2i\kappa}} \ket{\psi'} \\
    \ket{\omega_-} & = \frac{-r_2e^{-2i\kappa}}{r_1^2 - r_2^2 e^{-2i\kappa}}
    \ket{\psi} + \frac{r_1e^{-i\kappa}}{r_1^2 - r_2^2 e^{-2i\kappa}} \ket{\psi'}
\end{align*}

Therefore, we obtain the
three following projectors to distinguish between $\ket{\psi}$ and
$\ket{\psi'}$: $\{P_0=\ket{\omega_+}\bra{\omega_+}, P_1 =
\ket{\omega_-}\bra{\omega_-}, P_? = \mathbb{I} - P_0 - P_1\}$ with outcomes \0,
\1 and \?, respectively.
The outcome \0 corresponds to the output state being $\ket{\psi}$ and hence
implies that the circuit is (probably) fault-free; similarly, outcome \1 implies
that the $i$-th gate is probably faulty. Outcome \? is never observed if circuit
is fault-free or if the $i$-th gate is faulty; therefore, outcome \? immediately
signifies that the circuit has fault at some other gate. We capture the
measurement output using triplets containing probability of different outcomes
$(\p(0),\p(1),\p(?))$.
We call the combination of a $(C^0,C^i)$-separator input and a measurement
operator to distinguish between $C^0$ and $C^i$ as $Test(i)$.

The probability of error after one measurement would be at most
$\delta = |\braket{\psi}{\omega_-}|^2 = r_2^2 = \left(1-\sqrt{1-k^2}\right)/2$
which matches the
minimum probability of error in distinguishing $\ket{\psi}$ and
$\ket{\psi'}$ by any projective measurement ~\cite{helstrom}. This shows that
$Test(i)$ can optimally distinguish between a faulty and a fault-free $i$-th gate.
Table \ref{table:gate-separator-input} shows the
probability of error in detecting SMGF faults for some of the commonly used
quantum gates. The table demonstrates that for most gates, missing gate faults
can be easily detected.
If necessary, $\delta$ could be further reduced using standard techniques of
repeating a $Test$ and reporting the majority of measurement
outcomes --- $O(\frac{1}{\delta}\log\frac{1}{\epsilon})$ repetitions are
required to reduce error to $\epsilon$.

\begin{figure}[b]
    \begin{tikzpicture}[thick,scale=1, every node/.style={scale=1}]
\tikzset{
operator/.style = {draw,fill=white,minimum size=1.5em},
operator2/.style = {draw,fill=white,minimum height=2cm},
phase/.style = {draw,fill,shape=circle,minimum size=5pt,inner sep=0pt},
surround/.style = {fill=blue!10,thick,draw=black,rounded corners=2mm},
cross/.style={path picture={ 
\draw[thick,black](path picture bounding box.north) -- (path picture bounding box.south) (path picture bounding box.west) -- (path picture bounding box.east);
}},
crossx/.style={path picture={ 
\draw[thick,black,inner sep=0pt]
(path picture bounding box.south east) -- (path picture bounding box.north west) (path picture bounding box.south west) -- (path picture bounding box.north east);
}},
circlewc/.style={draw,circle,cross,minimum width=0.3 cm},
dotted/.style={draw,dotted,minimum size=1.7em},
}
\matrix[row sep=0.4cm, column sep=0.8cm] (circuit) { 
\node (q1) {};  
& \node[phase] (P11) {};
& \node[operator] (H21) {H};
&
& \node[] (U21) {};
&  \node[phase] (P12) {};
&  
\coordinate (end1); \\

\node (q2) {};    
& \node[phase] (P20) {};                                    
&
&
&
&\node[circlewc] (P23) {};
&\coordinate (end2);\\
\node (q3) {};    
& \node[circlewc] (P30) {};                                     
&\node[operator] (H31) {H}; 
&\node[](U31){};                                      
&
& 
&\coordinate (end3);\\
};

\begin{pgfonlayer}{background}
\draw[thick] (q1) -- (end1)  
(q2) -- (end2) 
(q3) -- (end3) 
  (P11) -- (P20) -- (P30) (P12) -- (P23);
 \node[operator2] at (U31){\rotatebox{90}{$R_y(\pi/6)$}};  
\node[operator2] at (U21){\rotatebox{90}{$R_z(\pi/16)$}};
\end{pgfonlayer}
\end{tikzpicture}
\caption{\label{fig:3qubitcnot}Benchmark circuit 3qubit-CNOT}
\end{figure}
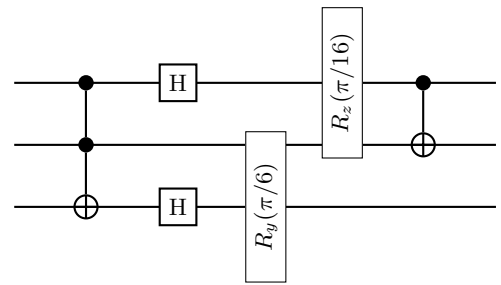

For deciding if $C$ is $C^0$ or $C^i$, we first compute the
two triplets $\mu^{ff}=(1-\delta, \delta, 0)$ for fault-free and $\mu^F=(\delta, 1-\delta, 0)$
for faulty circuit. Then we estimate the distribution of measurement outcomes
by running the circuit multiple times using the separator state as
input. Standard statistical techniques of
classification can be used to determine if the observed distribution is obtained from
$\mu^{ff}$ or $\mu^F$. The optimum number of samples
required is inversely proportional to the L1 distance of these distributions,
which in our case is equal to $(1-2\delta)$~\cite{BJV2004}. It is clear from
Table \ref{table:gate-separator-input} that just a single measurement can
determine if a particular Hadamard gate is missing.

\section{Detecting if a circuit is faulty}
\label{sec:algo-faulty-circuit}

Having discussed our solution to the problem of deciding whether a particular
gate in a quantum circuit is faulty or not, given that other gates are
fault-free, now we discuss the more general case where {\em any one
gate} in a circuit may be faulty. Our efficient diagnostic strategy uses a {\em
pre-processing} stage and a {\em circuit evaluation} stage.

\paragraph*{Pre-processing stage:} 
The pre-processing stage takes as input
a description of the circuit, along with each of the fault-free and faulty
operators. First, for each gate $G_i$, we construct the input state and
measurement operator for $Test(i)$. Then, we
construct a {\em diagnostic table} with $s$ rows and $(1+s)$ columns whose $(q,r)$-th cell
contains the triple $\pi(q,r)=(\Pr[\0], \Pr[\1], \Pr[\?])$ when $Test(q)$ is applied to
circuit $C^r$ --- the probabilities can be obtained by using any quantum circuit
simulator, such as QuIDDPro\cite{quiddpro}. Diagnostic tables for a benchmark circuit
{\em 3qubitcnot} (illustrated in Figure \ref{fig:3qubitcnot}) are given in the
Table \ref{table:fault-table-proj}.

\paragraph*{Circuit evaluation stage:} 
During circuit evaluation stage, we get an input circuit $C$ on which we can
apply any $Test$ from $\{Test(F1) \ldots  Test(Fs)\}$. Suppose we apply $Test(Fi)$ (for some chosen $i$)
enough number of times to create an output distribution $\hat{\pi}$. If
$C=C^j$, then $\hat{\pi}$ will be ``closest'' to the distribution $\pi(i,j)$.
What is required is therefore an efficient way to classify $C$ into the classes
$\{C^0 \ldots C^s\}$ by multiple applications of suitable Tests.
It is clear that $Test(j)$ will be able to identify between $C^0$ and $C^j$
since the L1 distance of $\pi(j,0)$ and $\pi(j,j)$ is at least $(1-2\delta)$
($\delta$ is a property of the $j$-th gate as explained earlier). However, in practice
it is possible that a particular $Test()$ is able to distinguish between more
than two classes of faults. The diagnostic table for the 3qubitcnot circuit
shows that most Tests are able to easily classify all $C^i$, except $C^5$ \& $C^0$ in some cases. Therefore,
Test(1) followed by Test(5), a few applications of each, 
suffices to diagnose all faults
-- less that 20 evaluations of $C$ which is a huge improvement compared to earlier works~\cite{paler1,paler2}.

\section{Conclusion}
In this letter we present a clear outline on how one should diagnose faults in
quantum circuits in the framework of ATPG.  Our explanation is mostly based on single gate faults, but the
techniques can be extended to certain types of multiple faults, though the computational costs
would naturally increase in such cases. We show how to isolate every type of
fault using very few test patterns compared to existing techniques. Our main contribution here is to
demonstrate that while studying faults in quantum circuits, one should consider
the properties of quantum circuits for choosing proper quantum input states as well as proper strategy of
measurement for efficient testing of circuits.


\begin{acknowledgments}
First author would like to thank the Centre of Excellence in Cryptography,
Indian Statistical Institute, Kolkata
for hosting him during which major portion of this work was accomplished. 
\end{acknowledgments}

\end{document}